\documentclass[final,1p,times]{elsarticle}

\usepackage{amssymb}

\usepackage{graphicx}
\usepackage{multirow}
\usepackage{dcolumn}
\usepackage{bm}
\usepackage{amsmath}
\usepackage{braket}
\usepackage{subfigure}

\journal{Physics Letters B}

\begin{document}

\begin{frontmatter}



\title{On the $\beta$-decay of $^{186}$Hg}



\author[a1,a2]{A.~Algora\corref{cor1} }
\ead{algora@ific.uv.es}
\cortext[cor1]{Corresponding authors}
\author[a3]{E.~Ganio\u{g}lu\corref{cor1} }
\ead{ganioglu@istanbul.edu.tr}
\author[a4]{P.~Sarriguren}
\author[a1,a5]{ V.~Guadilla}
\author[a6]{ L.~M.~Fraile}
\author[a1,a4]{ E.~N\'{a}cher}
\author[a1]{ B.~Rubio}
\author[a1]{ J.~L.~Tain}
\author[a1]{J. ~Agramunt}
\author[a8]{ W.~Gelletly}
\author[a4,a9]{J.~A.~Briz}
\author[a3]{R.~B.~Cakirli}
\author[a9]{ M.~Fallot}
\author[a1]{D.~Jord\'an}
\author[a2]{ Z.~Hal\'asz}
\author[a2]{ I.~Kuti}
\author[a1]{ A.~Montaner}
\author[a9]{ A.~Onillon}
\author[a1]{ S.~E.~A.~Orrigo}
\author[a1]{ A.~Perez~Cerdan}
\author[a8]{ S.~Rice}
\author[a6]{ V.~Vedia}
\author[a1]{ E.~Valencia}

\address[a1]{Instituto de F\'{\i}sica Corpuscular, CSIC-Univ. de Valencia, E-46071 Valencia, Spain}
\address[a2]{Institute of Nuclear Research (ATOMKI), P.O.Box 51, H-4001 Debrecen, Hungary}
\address[a3]{Department of Physics, Istanbul University, Istanbul 34134, Turkey}
\address[a4]{Instituto de Estructura de la Materia, CSIC, E-28006, Madrid, Spain}
\address[a5]{Faculty of Physics, University of Warsaw, 02-093, Warsaw, Poland}
\address[a6]{Univ. Complutense, Grupo de F\'{\i}sica Nuclear, CEI Moncloa, E-28040, Madrid, Spain}
\address[a8]{Department of Physics, University of Surrey, GU2 7XH, Guildford, UK}
\address[a9]{Subatech, IMT-Atlantique, Univ. de Nantes, CNRS-IN2P3, F-44307, Nantes, France}

\begin{abstract}
The Gamow-Teller strength distribution of the decay of $^{186}$Hg into $^{186}$Au has been determined for the first time using the total absorption gamma spectroscopy technique and has been compared with theoretical QRPA calculations using the SLy4 Skyrme force. 
The measured Gamow-Teller strength distribution and the half-life are described by mixing oblate and prolate configurations independently in the parent and daughter nuclei. The best description of the experimental beta strength is obtained with dominantly prolate components for both parent $^{186}$Hg and daughter $^{186}$Au. The approach also allowed us to determine an upper limit of the oblate component in the parent state. 
The complexity of the analysis required the development of a new approach in the analysis of the X-ray gated total absorption spectrum. 
\end{abstract}


\begin{highlights}
\item The Gamow-Teller strength distribution of the decay of $^{186}$Hg into $^{186}$Au has been determined for the first time using the total absorption gamma spectroscopy technique and has been compared with theoretical QRPA calculations using the SLy4 Skyrme force. 
The measured Gamow-Teller strength distribution and the half-life are described by mixing oblate and prolate configurations independently in the parent and daughter nuclei. The best description of the experimental beta strength is obtained with dominantly prolate components for both parent $^{186}$Hg and daughter $^{186}$Au. The approach also allowed us to determine an upper limit of the oblate component in the parent state, which also describes nicely the experimental beta strength and provides the best description of the half-life of the decay within this framework. 
\item The complexity of the analysis required the development of a new approach in the analysis of the X-ray gated total absorption spectrum. This approach 
can also be of particular interest for cases where the $\beta^+$ component of the $\beta$-decay can contaminate the X-ray gated spectra. 
\end{highlights}


\end{frontmatter}








The existence of eigenstates characterized by different intrinsic shapes in a particular nucleus can be considered a unique type of behaviour in finite many-body quantum systems \cite{shapecoexistence3}. This phenomenon, called shape coexistence, is essentially a quantum mechanical effect that appears very clearly in specific regions of the nuclide chart \cite{shapecoexistence3,shapecoexistence2,shapecoexistence1}. 
The appearance of co-existent structures in nuclei has been interpreted as the consequence of the competition of two opposing trends: the stabilizing effect of closed shells or subshells, that drives the nuclear system to sphericity and the residual interactions between protons and neutrons that drives the system to deformation. 
Theoretically it has been interpreted in the framework of the shell model or mean-field approaches.  
The coexisting structures can mix depending on their quantum mechanical properties, and the resulting states can be connected by transitions basically determined by their degree of mixing. 
The relevant states have been studied conventionally through electromagnetic probes, nuclear transfer reactions and $\alpha$-decay, but its possible impact in $\beta$-decay has only been studied in a few cases
 (for the most recent reviews see \cite{shapecoexistence3,focus1,focus2} and references therein). 



The region around the neutron-deficient Hg nuclei is considered to be a benchmark for studies related to shape transitions and shape effects. Its study has attracted significant attention in relation to measurements of the changes in mean-square charge radii ($\delta \langle r^2 \rangle$) \cite{shapecoexistence_Hg}.  
The $\delta \langle r^2\rangle $ measurements show a characteristic staggering for the Hg isotopic chain, that is not seen so clearly in any other isotopic chain in the nuclide chart \cite{Angeli}. The staggering was interpreted as a change in the ground state structure and consequently on the ground state shape around A=186
\cite{Frauendorf}. A recent work performed at CERN-ISOLDE \cite{Marsh2018} extended the study of the $\delta \langle r^2\rangle$ in the Hg nuclei down to A=179 and established firmly the limits of the staggering phenomenon. This study also confirmed the results of the earlier $\delta \langle r^2\rangle$ measurements. 

In this context, $\beta$-decay studies can offer an insight into nuclear shapes and related phenomena in particular cases. The idea, first introduced by I. Hamamoto and coworkers \cite{Hamamoto} and later developed by P. Sarriguren {\it et al.} \cite{Sarriguren}, 
is based on the dependence of the $\beta$ strength of the decay 
on the nuclear shape assumed for the parent state. Information on the deformation of the ground state of the decaying nucleus can be obtained from the comparison between experiment and theory in cases where the pattern of the theoretical strength distributions show a clear dependence on the shape \cite{Nacher,Poirier,Perez,Briz,Aguado}. For a proper determination of the experimental $\beta$ strength, these studies require the application of the total absorption gamma-ray spectroscopy (TAGS) technique, that provides $\beta$-decay data free from the Pandemonium systematic error \cite{Hardy}.

These studies have relied mainly on quasiparticle random-phase approximation (QRPA) calculations, 
where the deformation of the parent state and the deformation of the populated states in the daughter nucleus remain the same. For that reason the method was originally thought to be applicable mainly to cases where mixing in the ground state of the parent nucleus was assumed to be small. Thus in the case of the $\beta$-decay of $^{74}$Kr, where the theoretical description of the measured $\beta$ strength was relatively poor compared to the $^{76}$Sr case \cite{Nacher}, this was interpreted as evidence of a strongly mixed, ground state configuration in the parent nucleus \cite{Poirier}.





\begin{figure}
	\includegraphics[width=1.0\textwidth]{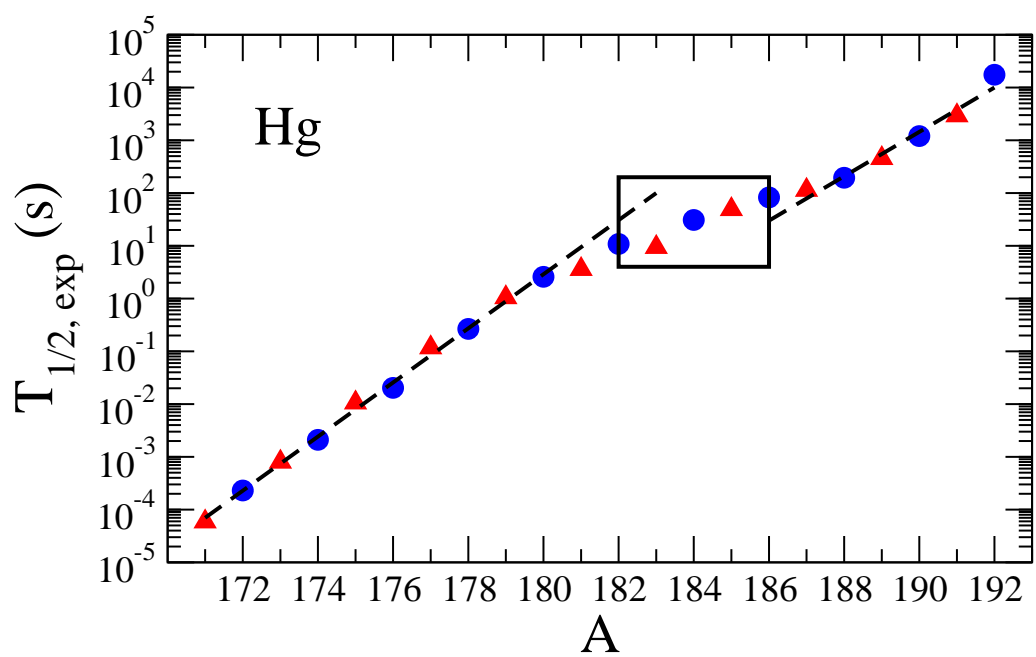}
	\caption{\label{fig:half_life_trend} Systematics of the $\beta$-decay half-lives of Hg isotopes. From the figure two trends are visible, and a transitional region between A=180-188. The lines are provided as a guide to the eye. Circles are used for even-even nuclei, and triangles for even-odd cases.}
\end{figure}

In this context, the $\beta$-decay of $^{186}$Hg can be seen as a one-off. 
From the pattern observed in the differences in charge-radii, the ground states of even-even Hg isotopes around $A=186$ are compatible with oblate shapes that are in general less common in the nuclide chart \cite{Alejandro}, while the odd-$A$ isotopes below $A=186$ are best described with prolate shapes \cite{Hilberath}. Similarly, the jump in the charge-radii differences observed in Au isotopes between $A=187$ and $A=186$ is a signature of a shape transition between an oblate shape in $^{187}$Au and a prolate shape in $^{186}$Au \cite{Kroner}.
Therefore, we face in this case an interesting problem, where a $\beta$-decay can connect partners with 
deformations that are assumed to be quite different, at least in their ground states. They are expected to be predominantly oblate in the parent $^{186}$Hg and predominantly prolate in the daughter $^{186}$Au.
The decay in these cases is expected to be supressed with respect to decays between partners with similar shapes. Actually, this suppression can be observed in the measured half-lives of Hg isotopes, where the trend changes when crossing the decay around $A=186$ (see Fig. \ref{fig:half_life_trend}). In this figure the trend observed in the half-lives of the heavier Hg isotopes ($A>186$) would correspond to decays where parent and daughter have similar shapes. The trend is broken in the mass region $181<A<186$, where the half-lives are somewhat larger, as expected from decays connecting partner nuclei with different shapes. A smooth trend is recovered in the decay of the lighter Hg isotopes.

Theoretical calculations \cite{Sarriguren_Hg,Boillos} show that in the $^{186}$Hg case there is a clear sensitivity of the $\beta$ strength of the decay to the shape of the parent nucleus independently of the force employed in the calculations, which justifies such study.
Even though the information that can be obtained from such a $\beta$-decay study is considered model dependent, it can be complementary to the results based on other techniques such as Coulomb excitation, and can provide 
independent information on the prolate or oblate character of the ground state of the parent nucleus, since it is based on a different probe.  This is of particular interest since a direct measurement of the electric quadrupole moment of the ground state of $^{186}$Hg is not possible. 
Furthermore, even in the most recent work by Bree {\it et al.} \cite{Bree} it was not possible to obtain information on the sign of the deformation for the $^{186}$Hg ground state. 

\begin{figure}
	\includegraphics[width=1.0\textwidth]{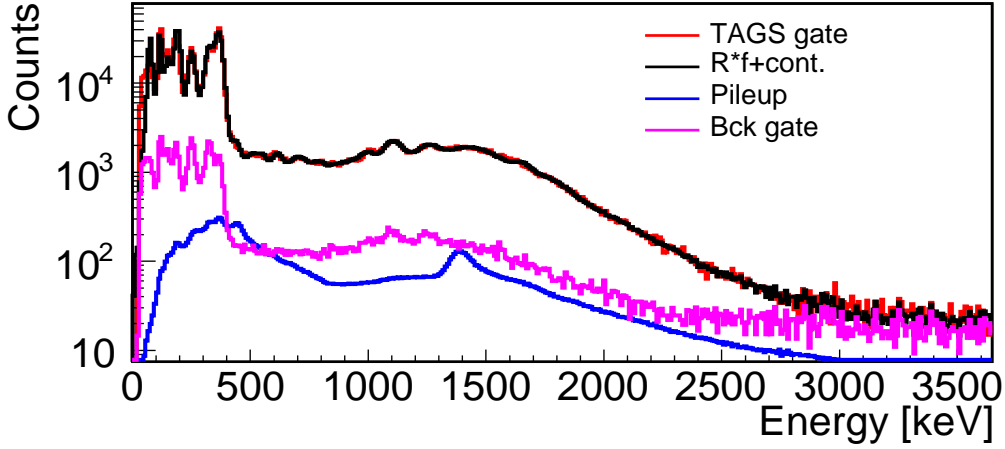}
	\caption{\label{fig:spectrum} Comparison of the analyzed TAGS X-ray gated spectrum (red) and reconstructed (black) spectrum after the analysis for the $^{186}$Hg decay. The reconstructed spectrum is calculated with the $\beta$-intensities obtained in the analysis and the corresponding response function of the spectrometer. The contributions of the different contaminants (pileup and background) are also presented (blue and pink, respectively).}
\end{figure}

\begin{figure}
	\includegraphics[width=1.0\textwidth]{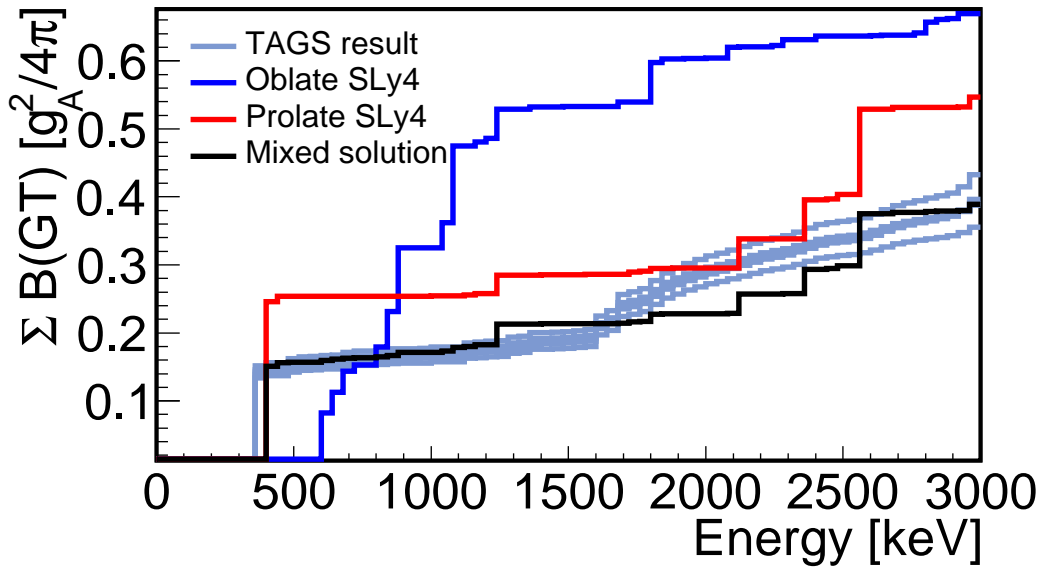}
	\caption{\label{fig:sly4} Accumulated $\beta$ strength deduced from the analysis as a function of the level energy compared with the theoretical calculations using SLy4 force \cite{Boillos} for the decay of $^{186}$Hg. The mixed solution corresponds to ($\lambda,\alpha$)=(0.46,0.46). For more details see the text.}
\end{figure}

This article summarizes the results of the first $\beta$-decay investigation of $^{186}$Hg using the total absorption technique. 
The study was challenging both in terms of the analysis and the interpretation, and further details will be given elsewhere \cite{Long_Ela}.
On one hand, the measured total absorption spectra for the decay is quite exceptional compared to our previous experience, 
since the most important features of the decay spectrum are concentrated at very low excitation energy in a "comb" like structure (see Fig. \ref{fig:spectrum}).  
The study of this decay required the development of a new approach in the analysis because of the sizable  conversion electron coefficients of the gamma transitions that de-excite the strongly populated $1^{+}$ state at 363 keV in $^{186}$Au and to the penetration and summing of the X-rays in the total absorption spectrometer employed. On the other hand, the straightforward interpretation of the experimental data, 
based on the direct comparison of the deduced $\beta$ strength with the theoretical calculations, seems at odds with assumed facts in the region and was further examined (see Fig. \ref{fig:sly4}). 
In order to obtain a consistent picture of the whole phenomenology, an additional assumption, that both parent and daughter states can be mixed in different degrees was required. The result from this study shows that it is possible to find a given mixing of oblate and prolate configurations 
for both parent and daughter nuclei that is able to reproduce nicely the measured $\beta$ strength. 
The theoretical analysis presented here for the first time in the framework of QRPA calculations should be considered in cases where parent and daughter nuclei are expected to exhibit different degrees of mixing.


The $\beta$-decay of $^{186}$Hg was studied at CERN-ISOLDE. In this experiment Hg isotopes were produced by bombarding a 50 gcm$^{-2}$ UC$_{x}$ target with a beam of  1.4 GeV protons delivered by the Proton Synchrotron Booster. In the measurements, the RILIS laser ionisation source ~\cite{Rilis} was used to ionise selectively Hg isotopes before separation in mass ($A$) with the General Purpose Separator (GPS). Then, the mass separated beam was transported to the total absorption spectrometer (TAS) {\it Lucrecia}, where it was implanted outside the spectrometer in the magnetic tape of a tape transport system. The tape was used to move the accumulated activity to the centre of the spectrometer in collection and measuring cycles that were determined by the half-life of the isotope of interest. In the case 
of the $^{186}$Hg decay study ($T_{1/2}$=1.38(6) min \cite{Baglin}), the measuring cycle was optimized to 2 min, in order to reduce the impact of the daughter activity in the measurements. 

The TAS detector \textit{Lucrecia}, is made of a  cylindrically shaped NaI(Tl) mono-crystal with 38 cm diameter and 38 cm length.
The total efficiency of this setup has been estimated using Monte Carlo (MC) simulations to be 90 \% for mono-energetic gamma rays in the range of 300-3000 keV, which gives an approximately 99  \% efficiency for gamma cascades of more than one gamma ray. In this setup the beam pipe is inserted in a hole of $\phi =7.5$ cm perpendicular to the symmetry axis of the detector to reach the optimal measurement position
inside the spectrometer.  
Opposite to the beam pipe end cap, ancillary detectors are placed 
(for a schematic figure of the setup see Ref. \cite{Perez}). 
As in earlier measurements of the $\beta$-decay of neutron deficient cases, 
 two ancillary detectors were used: a germanium telescope, that is composed of a Ge planar and a Ge coaxial detector, together with a thin plastic beta detector. 
The ancillary detectors allow us to tag the electron capture (EC) or the $\beta^+$ component of the decay by requesting coincidences of the TAGS spectrum with the X-rays detected in the planar detector (EC component) or with the $\beta$-particles detected in the $\beta$ detector ($\beta^+$ component). More details on the \textit{Lucrecia} setup can be found in \cite{Perez,Briz,Rubio}. 
In this kind of study, the TAGS spectrum generated in coincidence with the X-rays emitted in the EC process is preferred for the 
analysis, since the coincidence is element selective and provides a very clean decay spectrum. 

For the analysis of the total absorption spectrum, the  \mbox{$d_i = \sum_{j=0}^{j_{max}}R_{ij}(B)f_j + c_i$ $(i=1,.., i_{max})$} matrix equation has to be solved. Here $d_i$ represents the content of bin $i$ in the TAGS spectrum, $R_{ij}(B)$ is the response matrix of the setup and represents the probability that a decay that feeds level $j$ in the daughter nucleus gives a count in bin $i$ of the TAGS spectrum ($d_i$),  $f_j$ is the $\beta$ feeding to the level $j$, that has to be determined and $c_i$ represents the contribution of possible contaminants to the contents of bin $i$. The TAGS spectrum $d$, used in the analysis, was generated by putting a condition on the X-rays of Au, specifically coincidences with the $K_{\alpha 1}$  and  $K_{\beta 2}$ lines.  The contributions of the X-ray background and pileup to the spectrum $d$ were also taken into account. The calculation of the response function $R_{ij}(B)$ requires the knowledge of the branching ratio matrix $B$ of the levels in the daughter nucleus. In our conventional analysis this matrix is first calculated using the information available from high-resolution studies up to a certain threshold energy, in this particular case up to 600 keV.  Above that energy threshold and up to the decay $Q$ value (3176(24) keV \cite{Wang} ) the  statistical model is used to generate the gamma decay branches of the levels.  The statistical model is based on a Back-Shifted Fermi Gas Model level density function \cite{Dilg1973} and gamma strength functions \cite{Capote2009} of E1, M1, and E2 character. The level density function parameters were obtained from fits to the data available in \cite{Goriely2001,Demetriou2001,Capote2009}. Once the branching ratio matrix is defined, it is possible to calculate the $R_{ij}$ using MC techniques and solve the equation using an appropriate algorithm \cite{Tai07a,Tai07b}.

The first analysis attempts, that followed the conventional procedure of analysis, were not able to reproduce the X-ray gated TAGS spectrum of the decay of $^{186}$Hg. Additional peaks appear in the X-ray gated experimental TAGS spectrum, that are not accounted for in our conventional response function. The reason lays in the way of calculating the response function where X-rays were not generated and not included in the gamma-ray response. This is conventionally considered unnecessary, because of the low energy of the X-rays involved in most of the $\beta$-decay cases studied and their absorption by the layers of dead material in the total absorption detectors and materials around the sources. 
According to the high-resolution study of Porquet {\it et al.} \cite{Porquet}, the state that receives most of the feeding in the $\beta$-decay of  $^{186}$Hg is the $1^{+}$ state at 363.6 keV excitation energy in $^{186}$Au. The total absorption peak associated with this state is a clearly dominant feature in the TAGS spectrum (see Fig. \ref{fig:spectrum}). The 363.6 keV state de-excites by a gamma cascade of  two gamma rays (112.1 keV (E1) - 251.5 keV (M1)). Both gamma rays have sizable internal conversion coefficients  ($\alpha_{tot}$=0.315 for the 112.1 keV and $\alpha_{tot}$=0.558 for 251.5 keV transitions respectively) \cite{Baglin}, which implies that X-rays are generated when electron conversion occurs, 
independently of whether the 363.6 keV state was populated in the $\beta$-decay by EC (with the consequent generation of X-rays) or by $\beta^+$ transitions.

These additional X-rays, and their large energies ($K_{\alpha1}$(Au)=68.8 keV) caused the difficulties in the conventional analysis. Assuming a pure EC decay to the 363.6 keV state for simplicity, in the case of electron conversion of any gamma transition in the cascade, the resulting X-ray from internal conversion can be summed with the unconverted gamma transition of the cascade, thus generating additional peaks in the TAGS spectrum.  Thus for example we will have peaks at 112.1 keV plus an X-ray from conversion of the 251.5 keV transition, and 251.5 keV plus an X-ray from conversion of the 112.1 keV transition. In addition, X-rays generated by internal conversion can lead to a $\beta^+$ contamination in the EC spectrum (defined experimentally by the X-ray coincidence, since the 363.6 keV state can be populated also via $\beta^+$ decay and we can detect one of the X-rays generated from the electron conversion of any of the gamma lines 
de-exciting the level in the planar detector). To address the complexity of the problem (generation of X-rays because of the EC/$\beta^+$ competition depending on the excitation of the level, the generation of X-rays because of conversion and the summing of gamma transitions with X-rays) a new way of calculating the response function was developed. The method will be described in full detail in forthcoming publications \cite{Long_Ela, NIM_Ale}, but essentially what is done is to exploit fully the tools provided by the radioactive decay package of the GEANT4 code \cite{Agostinelli}. The response function for each level in the daughter nucleus is obtained by simulating a $\beta$-decay (EC + $\beta^+$) using an artificially generated ``$\beta$-decay level scheme" that assigns $\beta$-feeding to the level for which the response is calculated only, and considering that the level (populated in the $\beta$-decay) de-excites with a branching ratio matrix $B$ calculated by us. The branching ratio matrix of the level in the daughter is determined according to our conventional method, by combining the known branching ratio matrix from high-resolution studies for the low-lying levels with the added part at higher excitation energy from the statistical model in bins of 40 keV. The response to the level is obtained finally by collecting the obtained TAGS spectrum in the GEANT4 simulation in coincidence with the X-rays detected in the planar detector, in exactly the same conditions as in the experiment. By using this method we employ the tools provided by GEANT4 for the generation of X-rays in all the processes and the calculation of the EC/$\beta^+$ ratio for each level. The final response ($R_{ij}(B)$) matrix is obtained by combining the different responses obtained for each possible level populated in the decay up to the Q value.

The analysis employing the new calculation method of the response function, provided a nice reproduction of the X-ray gated TAGS spectrum (see Fig. \ref{fig:spectrum}). From the feeding distribution obtained the $\beta$ strength of the decay is deduced, which is compared with QRPA theoretical calculations using the SLy4 force in Fig. \ref{fig:sly4} \cite{Boillos}. The error bands of the experimental strength distribution are determined by the possible solutions of the TAGS inverse problem that reproduce reasonably well the experimental spectrum and by the errors of the $Q$ value and the $T_{1/2}$ of the decay used in the calculation of the strength  ($Q_{EC}$=3176(24) keV \cite{Wang}, $T_{1/2}$=1.38(6) \cite{Baglin}). 

An interpretation based on the direct comparison between the experimental strength and the calculations performed with pure shapes suggests clearly that the shape of $^{186}$Hg would be prolate in its ground state (see the red line in Fig. \ref{fig:sly4}),
but this result is in contradiction with the interpretation of the trend of $\delta \langle r^2 \rangle $ measurements around $^{186}$Hg, which has been explained as a rapid shape change from an oblate system ($^{186}$Hg) to a more deformed odd-A system in $^{185}$Hg. For that reason, the straightforward interpretation based on the direct comparison of the experiment with the results obtained from pure shapes was revised, and more complex deformation scenarios were considered. A simple mixing scenario of the parent (and consequently daughter), such as the one used to explain the $^{74}$Kr case was also discarded, since the variation of $\delta\langle r^2\rangle$ around $^{186}$Au was interpreted as a sign of $^{186}$Au being a dominantly  prolate nucleus \cite{Wallmeroth}.

\begin{figure}[!htbp]
        \includegraphics[width=1.0\textwidth]{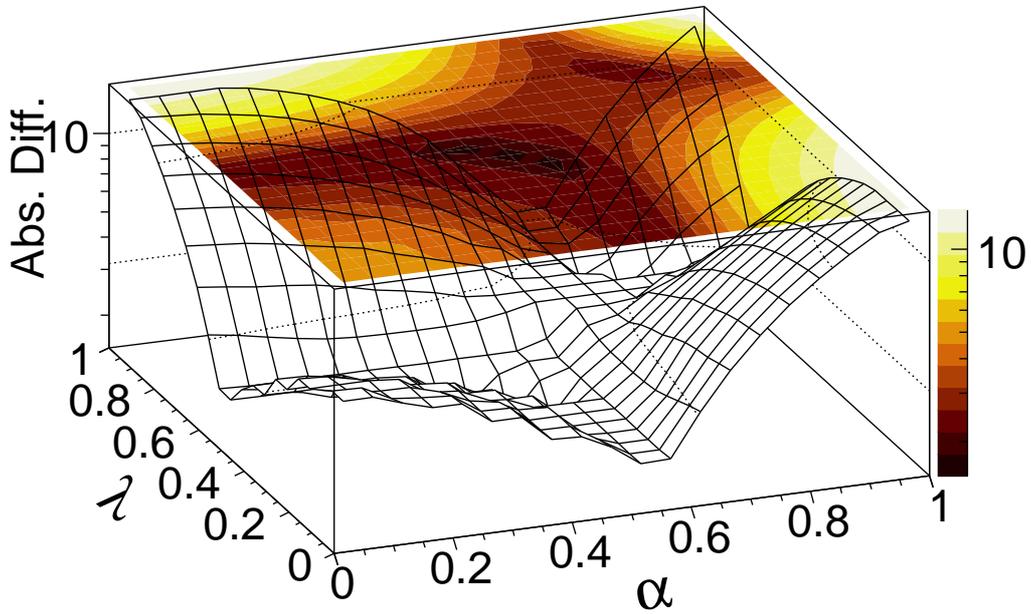}
	\caption{\label{fig:surface} Contour plot of the absolute value of the $GT_{exp}-GT_{mix}$ differences in the $\lambda$, $\alpha$ surface for the SLy4 force.        }
\end{figure}

Therefore, a more complex mixing scenario has to be necessarily considered in which the parent and daughter nuclei may both exhibit different degrees of mixing of the oblate and prolate components that correspond to the prolate ($\beta$=0.26) and oblate ($\beta$=-0.18) minima used in the QRPA calculations. 
This is carried out by the simple model given by Eqs. \ref{eq:1}, where parent ($\ket{\psi}_p$) and daughter ($\ket{\phi}_d$) states are described by the mixing of two shapes, oblate and prolate, with weights characterized by $\lambda$ and $\alpha$, respectively. The idea is to explore whether we can find independent combinations of prolate and oblate components, for both parent and daughter wavefunctions, that are able to reproduce optimally the experimental GT strength. 

\begin{figure}
	\includegraphics[width=1.0\textwidth]{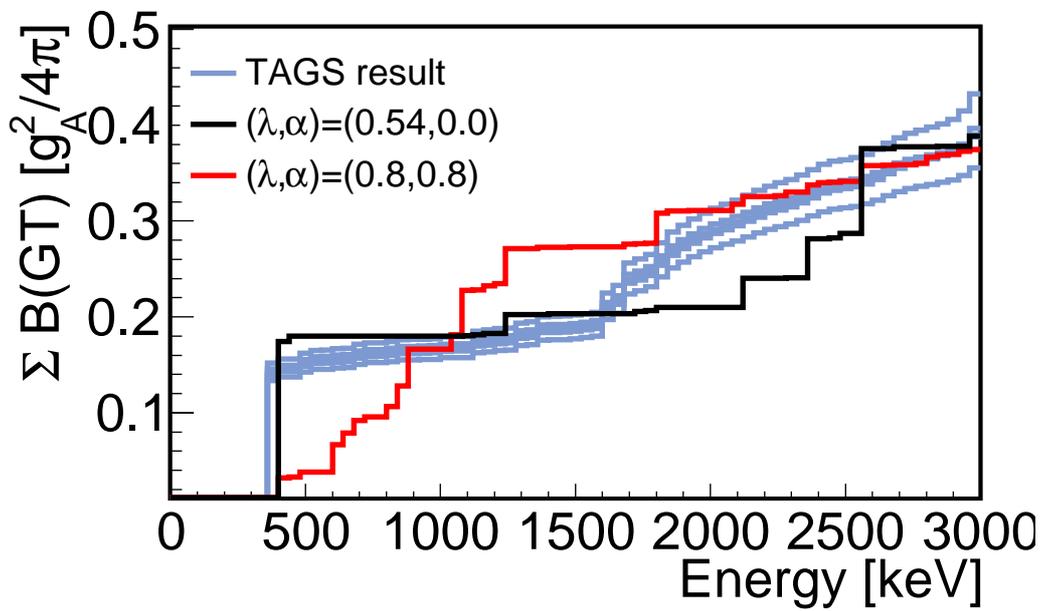}
	\caption{\label{fig:sly4_2} Accumulated $\beta$ strength deduced from the analysis as a function of the level energy compared with $GT_{mix}$ with ($\lambda,\alpha$)=(0.54,0.0) (black line) and (0.8,0.8) (red line).}
 \end{figure}


\begin{equation}
\label{eq:1} 
\begin{split}
\ket{\psi}_p &= \lambda\ket{\rm Oblate}_p+\sqrt{1-\lambda^2}\ket{\rm Prolate}_p \\
\ket{\phi}_d &= \alpha\ket{\rm Oblate}_d+\sqrt{1-\alpha^2}\ket{\rm Prolate}_d \\
GT_{mix}  &= \lambda^2 \alpha^2 GT_{Obl.} + (1-\lambda^2)(1-\alpha^2) GT_{Prol.} + C.T. \\
\end{split}
\end{equation}

It is worth emphasizing here what is meant by $GT_{Obl.}$ and $GT_{Prol.}$ in the formula. $GT_{Obl.}$ stands for the strength corresponding to the GT transition of the pure oblate ground state wave function of $^{186}$Hg to the oblate states in $^{186}$Au and similarly for $GT_{Prol.}$ in the prolate case. The consideration of the mixing scenario makes sense in this  particular case since the predicted oblate and prolate energy minima  for both $^{186}$Hg and $^{186}$Au are very similar. This is not only true for the calculations presented here, but also for the results from Hartree-Fock-Bogoliubov model calculations based on the Gogny force \cite{CEA_deformations}. In Eqs. \ref{eq:1}, C.T. stands for small cross terms that involve phases of the different components \footnote{In the studied case the C.T amounts to maximum 5 \% of the accumulated strength}. $GT_{mix}$ is the GT obtained from the assumed mixing of parent and daughter states.

As a first step we fitted the experimental $GT_{exp}$ (experimental GT strength) with the $GT_{mix}$ function defined in Eqs. \ref{eq:1} to find the $(\lambda,\alpha)$ parameters that best reproduce the data. The best fit of the experimental accumulated stregth was obtained with $\lambda$=$\alpha$=0.46, which implies that both parent and daughter nuclei have a dominantly prolate component in their ground states, as well as in the excited states of $^{186}$Au. The obtained $GT_{mix}$ is presented in Fig. \ref{fig:sly4} with a black line. Please note that in this model we are assuming that all states in the daughter nucleus have the same degree of mixing. This approximation can be considered a natural extension of the one employed in standard QRPA calculations where the shape of the parent nucleus and all the populated states in the daughter nucleus  are the same.

To get a deeper insight into possible additional mixing scenarios, a mesh of  $\alpha$ and $\lambda$ values was used to look for minima in the surfaces of the absolute value of the differences between $GT_{mix}$ and $GT_{exp}$
depending on the ($\alpha$,$\lambda$) coordinates.  
 
The GT-strength difference surface for the SLy4 force shows clearly two well defined minima at $(\lambda,\alpha)=(0.46,0.46)$ and $(0.8,0.8)$ (see Fig \ref{fig:surface}). The two minima are positioned in the diagonal of the $(\lambda,\alpha)$ plane, defining two valleys of quadrant shape.
The shape of the valleys is a consequence of the symmetrical dependence of the $GT_{mix}$ function on the $\lambda$ and $\alpha$ parameters. The deepest minimum at $(0.46,0.46)$ corresponds to the values already obtained from the simple function fit and describes very nicely the pattern of the accumulated strength as seen in Fig. \ref{fig:sly4}. The second minimum at $(0.8,0.8)$ describes well the accumulated value of the strength at $E_{exc}$=3.0 MeV and the pattern of the accumulated strength in the excitation energy interval of 2.0-3.0 MeV, but fails to describe well the pattern in the 0.4-1.8 MeV excitation energy interval in the daughter (see red line in Fig. \ref{fig:sly4_2}). 

The quadrant valley of solutions that contain the deepest minimum and consequently describes better the pattern of the energy dependence of the accumulated strength can be used to determine the maximum oblate content of the ground state of $^{186}$Hg that describes properly the experimental $\beta$-strength pattern. From the surface this value is obtained when $\alpha$ $\sim$ 0.0, and corresponds to a $\lambda$=0.54. This value is equivalent to an oblate content of the parent wave function of 29 $\%$ and correspondingly a 71 $\%$ prolate content. This solution provides a description of the pattern of the strength comparable in quality in terms of the $\chi^2$ with the one associated with the lowest minimum $(0.46,0.46)$, and provides the best description of the half-life (see Table \ref{tab:1} and Fig. \ref{fig:sly4_2}). It is worth mentioning that the description provided by the dominant oblate-oblate $(0.8,0.8)$ solution has a very poor $\chi^2$, compared to both $(0.46,0.46)$ and $(0.54,0.0)$ and fails to describe the decay to the predominant state at 363 keV in $^{186}$Au.

Summarizing, in this article we have presented for the first time the measurement of the $\beta$-decay of $^{186}$Hg using the total absorption technique. The complexity of the case required the development of a new approach in the analysis to handle the penetration and summing of the X-rays in the total absorption spectrometer that can also be of particular interest for cases where the $\beta^+$ component of the $\beta$-decay can contaminate the X-ray gated spectra. 
A double mixing scenario has been considered for the first time to interpret the results in the framework of QRPA calculations. 
From the analysis of the accumulated GT strength assuming a double mixing scenario, clearly the best description of the pattern of the strength is obtained with both parent and daugther nucleus having a dominantly prolate content (1-$\lambda^2$=0.78, at ($\lambda$,  $\alpha$)=(0.46,0.46)). The study of the minimum valleys of the ($\lambda$,  $\alpha$) surface allowed us to determine the maximum oblate component of $^{186}$Hg in its ground state, that also reproduces nicely the pattern of the accumulated GT-strength and provides the best description of the half-life of the decay (see Table \ref{tab:1}) in this framework. This corresponds to the ($\lambda$,  $\alpha$)=(0.54,0.0) values and  to an oblate content of  $\lambda^2$=0.29. These solutions (($\lambda$,  $\alpha$)=(0.46,0.46) and ($\lambda$,  $\alpha$)=(0.54,0.0)) further emphasize what is already apparent from Figs. \ref{fig:half_life_trend} and \ref{fig:sly4}, that $^{186}$Hg is located in a transitional region where the shape in the Hg isotopes is changing and the beta transition strengths might be dictated by a fraction of the wave function that determines the overlap of parent and daugher wave functions in the case of different parent and daughter shapes. 
Although model dependent, the results presented here represent an alternative
confirmation of the mixed character of the ground state of $^{186}$Hg.

\begin{table}
     \begin{tabular}{| c | c | c | c |}
     \hline
       & Shape    & T$_{1/2}$ [s]  & $(\lambda,\alpha)$\\ \hline
       & prolate   & 60.0           &   (0.0,0.0)            \\
      & oblate    &  47.9          &  (1.0,1.0)            \\ 
 QRPA-SLy4 & mixed    &  89.1          &   (0.46,0.46)             \\
  model    & mixed    &  84.5          &   (0.54,0.0)             \\
       & mixed    &  93.4          &   (0.80,0.80)             \\\hline
Experiment&               &  82.3(36)    &                \\  \hline
\end{tabular}
\caption{Comparison of the experimental value of the T$_{1/2}$ with the results of the calculations assuming different shape scenarios.
The corresponding $(\lambda,\alpha)$ coordinates are provided in the last column. Note that the best description of the ground state half-life is provided by the $(0.54,0.0)$ mixed scenario, which maximises the oblate content of the parent ground state (see the text for details). }
\label{tab:1}
\end{table}

This work was supported by Spanish Ministerio de Econom\'{\i}a y Competitividad under grants FPA2011-24553, FPA2014-52823-C2-1-P, FPA2017-83946-C2-1-P, Ministerio de Ciencia e Innovacion grant PID2019-104714GB-C21, the program Severo Ochoa (SEV-2014-0398), and grants RTI 2018-098868-B-100 and ENSAR (grant 262010). S. E. A. O thanks the support of CPAN Consolider-Ingenio 2010 Programme CSD2007-00042 grant. E. G. acknowledges support from TUBITAK 2219 Abroad Research Fellowship Programme. R. B. C. acknowledges support by the Max-Planck-Partner group. Support from the technical staff and engineers of ISOLDE-CERN is acknowledged. W.G. acknowledges the support of STFC(UK) council grant ST/P005314/1. 
The help of Karl Johnston in the preparation of the $^{24}$Na source is also acknowledged. Enlightening discussions with Peter O. Hess, Thomas E. Cocolios and Sophie Peru are also acknowledged. 
This work was also supported by the National Research, Development and Innovation Fund of Hungary, financed under the K18 funding scheme with Projects No. K 128729 and NN128072. P. S. acknowledges support from MCI/AEI/FEDER,UE (Spain) under grant PGC2018-093636-B-I00. 

\bibliographystyle{elsarticle-num}

\end{document}